\documentclass[12pt]{aastex631}

\setcitestyle{numbers,square}

\usepackage{rotating}
\usepackage{amsmath}

\newcommand\n            {\noindent}
\newcommand\si           {\smallskip\indent}
\newcommand\bn           {\bigskip\noindent}
\newcommand\sn           {\smallskip\noindent}
\newcommand\cl           {\centerline}
\newcommand\ve           {\vfill\eject}

\newcommand\arcspt       {{$\buildrel{\prime\prime}\over .$}}
\newcommand\degree       {{\ifmmode^\circ\else$^\circ$\fi}} 
\newcommand\arcm         {{\ifmmode {'\ }\else$'     $\fi}} 
\newcommand\arcs         {{\ifmmode{''\ }\else$''    $\fi}} 
\newcommand{\bul}        {$\bullet$\ }

\newcommand\cge          {{$\gtrsim$}}
\newcommand\cle          {{$\lesssim$}}
\newcommand\degsq        {{deg$^{2}$}}

\newcommand\eg           {{\it e.g.},}
\newcommand\ie           {{\it i.e.}, }

\newcommand\kms          {{km\ s$^{-1}$}}

\newcommand\mum          {{\micron}}

\newcommand\nWsqmsr      {{nW\ m$^{-2}$\ sr$^{-1}$}}

\def\ltsima{$\; \buildrel < \over \sim \;$}
\def\lsim{\lower.5ex\hbox{\ltsima}}
\def\gtsima{$\; \buildrel > \over \sim \;$}
\def\gsim{\lower.5ex\hbox{\gtsima}}

\newlength{\txw}
\newlength{\txh}

\begin{document}

\vspace*{-1.300cm}
\title{We Must Preserve Hubble given its Unique Complementarity to Webb,
Roman, and Euclid}

%%%

\author[0000-0001-8156-6281]{Rogier A. Windhorst}%%% Rogier.Windhorst@gmail.com
\affiliation{School of Earth and Space Exploration, Arizona State University,
Tempe, AZ 85287-6004, USA}

\author[0009-0007-0782-0721]{Gibson B.\ Bowling} %%% gbbowlin@asu.edu
\affiliation{School of Earth and Space Exploration, Arizona State University,
Tempe, AZ 85287-6004, USA}

\author[0000-0002-6917-0214]{Kevin S. Croker}
\affiliation{School of Earth and Space Exploration, Arizona State University,
Tempe, AZ 85287-6004, USA}
%%% \affiliation{Department of Physics and Astronomy, University of Hawai`i at 
%%% M\=anoa, 2505 Correa Rd., Honolulu, HI, 96822}

\author[0000-0003-1625-8009]{Brenda Frye} %%% brendafrye@gmail.com
\affiliation{Steward Observatory, University of Arizona, 933 N Cherry Ave,
Tucson, AZ, 85721-0009, USA}

\author[0000-0001-6650-2853]{Timothy Carleton} %%% tmcarlet@asu.edu
\affiliation{School of Earth and Space Exploration, Arizona State University,
Tempe, AZ 85287-6004, USA}

\author[0000-0003-3329-1337]{Seth H. Cohen} %%% seth.cohen@asu.edu
\affiliation{School of Earth and Space Exploration, Arizona State University,
Tempe, AZ 85287-6004, USA}

\author[0000-0003-1268-5230]{Rolf A. Jansen} %%% rolfjansen.work@gmail.com 
\affiliation{School of Earth and Space Exploration, Arizona State University,
Tempe, AZ 85287-6004, USA}

\author[0000-0002-0648-1699]{Brent M. Smith} %%% Brent Smith <Brent.Smith.1@asu.edu>
\affiliation{School of Earth and Space Exploration, Arizona State University,
Tempe, AZ 85287-6004, USA}

\author[0000-0002-7265-7920]{Jake Summers} %%% jakesummers7200@gmail.com
\affiliation{School of Earth and Space Exploration, Arizona State University,
Tempe, AZ 85287-6004, USA}

\author[0000-0001-6265-0541]{Jessica M. Berkheimer} %%% jberkhei@asu.edu
\affiliation{School of Earth and Space Exploration, Arizona State University,
Tempe, AZ 85287-6004, USA}

\author[0000-0002-2099-639X]{Delondrae D. Carter} %%% ddcarte3@asu.edu 
\affiliation{School of Earth and Space Exploration, Arizona State University,
Tempe, AZ 85287-6004, USA}

\author[0000-0002-9984-4937]{Rachel Honor} %%% rchonor@asu.edu
\affiliation{School of Earth and Space Exploration, Arizona State University,
Tempe, AZ 85287-6004, USA}

\author[0000-0003-3351-0878]{Rosalia O'Brien} %%% obrienr2434@yahoo.com
\affiliation{School of Earth and Space Exploration, Arizona State University,
Tempe, AZ 85287-6004, USA}

\author[0000-0002-6150-833X]{Rafael {Ortiz~III}} %%% rortizii@asu.edu
\affiliation{School of Earth and Space Exploration, Arizona State University,
Tempe, AZ 85287-6004, USA}

\author[0000-0003-1949-7638]{Christopher J. Conselice} %%% conselice@gmail.com
\affiliation{Jodrell Bank Centre for Astrophysics, Alan Turing Building,
University of Manchester, Oxford Road, Manchester M13 9PL, UK}

\author[0000-0001-9065-3926]{Jose M. Diego} %%% chemadiegor@gmail.com
\affiliation{Instituto de F\'isica de Cantabria (CSIC-UC). Avenida. Los Castros
s/n. 39005 Santander, Spain}

\author[0000-0001-9491-7327]{Simon P. Driver} %%% Simon.Driver@icrar.org
\affiliation{
%%% International Centre for Radio Astronomy Research (ICRAR) and the
%%% International Space Centre (ISC), 
The University of Western Australia, M468,
35 Stirling Highway, Crawley, WA 6009, Australia}

\author[0000-0002-4884-6756]{Benne W. Holwerda} %%% benne.holwerda@louisville.edu
\affiliation{Department of Physics and Astronomy, University of Louisville,
Louisville KY 40292, USA} %%% 102 Natural Science Building

\author[0000-0001-9052-9837]{Scott A. Tompkins} %%% stompkins7192@gmail.com
\affiliation{
%%% International Centre for Radio Astronomy Research (ICRAR) and the
%%% International Space Centre (ISC), 
The University of Western Australia, M468,
35 Stirling Highway, Crawley, WA 6009, Australia}

\author[0000-0001-7592-7714]{Haojing Yan} %%% yanhaojing@gmail.com
\affiliation{Department of Physics and Astronomy, University of Missouri,
Columbia, MO 65211, USA}

%%% \email{Rogier.Windhorst@asu.edu}

\begin{abstract} We present compelling arguments --- focusing on galaxy science
--- for preserving the main imagers and operational modes of the Hubble Space
Telescope (HST) for as long as is technically feasible, to assure maximum
complementarity to the James Webb Space Telescope (JWST), Roman, and Euclid.
HST was designed to work well over the 0.1--1.6 \mum\ wavelength range, and
its unique UV--optical performance has fundamentally contributed to our
understanding of galaxy assembly and the Cosmic Star Formation History (CSFH).
While star-formation started at redshifts $z \gtrsim 10$, when the universe was
less than 500 Myr old, the CSFH did not peak until $z \simeq 1.9$ (\ie about 10
Gyr ago), and has steadily declined since that time. Hence, at least half of
all stars in the universe formed {\it in the last 10 Gyrs} where {\bf HST provides
its unique rest-frame UV view of unobscured young, massive stars tracing
cosmic star-formation, as well as unobscured Active Galactic Nuclei (AGN).} HST
thus uniquely probes (unobscured) young, hot, massive stars and AGN in
galaxies, while JWST, Euclid and Roman reveal more advanced stages of older
stellar populations, as well as relatively short-lived phases where galaxies
produce and shed a lot of dust from intense star-formation, dusty AGN, and the
very high redshift universe ($z \gtrsim 10$) not accessible by HST. {\bf HST is
thus \emph{highly complementary} to these other facilities, all of which took
decades to build to ensure decades of operation. To maximize return on
investment in these facilities, ways will need to be found to operate HST
imaging instruments in all relevant modes for as long as possible into the JWST
and Roman missions.} 
\end{abstract}

\vspace*{-0.800cm}
\n \section{INTRODUCTION: IMPACT OF HST IN THE LAST 32 YEARS} 

\sn The Hubble Space Telescope (HST) was designed in the 1960s and 1970s to
observe very faint objects at UV to near-IR wavelengths above the Earth's
atmosphere \citep[\eg][]{Smith1993}. HST's ability to observe outside the
Earth's atmosphere has resulted in very significant gains over ground-based
telescopes in four main areas, namely the ability to: (1) observe in the vacuum
ultraviolet; (2) observe with very stable, repeatable, and narrow Point-Spread
Functions (PSFs); (3) observe against very dark foregrounds and backgrounds;
and (4) perform precision (point-source) photometry at (very) high
time-resolution, but also on timescales of minutes to decades. As of April 24,
2026, HST has been in orbit for over 36 years. After successful correction of
the spherical aberration in its primary mirror in December 1993, HST has
produced an unprecedented wealth of high-quality data that has fundamentally
changed our understanding of the Universe, and of galaxy assembly and Super
Massive Black Hole (SMBH) growth in particular. The HST Archive presently
contains well over 3 million exposures from both its imagers and spectrographs. 

\si \emph{Over the last 32 years, HST has become the most successful science
mission ever undertaken.} From the STScI EPO office, we obtained the following
data to support this bold statement. HST had $\gtrsim 1200$ science press
releases since 1990, each with of order $\gtrsim 400$ million readers (or
impressions) worldwide. That is, HST press releases have resulted in
approximately $\sim480\times10^{9}$ reads (or impressions) worldwide. Thus, at
about $8\times10^{9}$ humans on Earth (assuming ages somewhat larger than
HST's), \emph{on average} each human would have read $\gtrsim 60$ HST
stories, and some smaller fraction would have read many more during their
lifetimes. HST is \emph{the most publicized} space astrophysics mission in NASA
history. HST has yielded on average $\gtrsim~$500--1300 refereed papers per
year by the astronomical community since 1990, counting $\gtrsim 50,300$ HST
related papers on 
\href{https://ui.adsabs.harvard.edu/search/fq=%7B!type%3Daqp%20v%3D%24fq_database%7D&fq_database=database%3A%20(astronomy%20OR%20physics%20OR%20general)&q=pubdate%3A%5B1990-01%20TO%209999-12%5D%20%20abs%3A(HST)&sort=date%20desc%2C%20bibcode%20desc&p_=0}
{ADS} as of May 2026 (with ``HST'' in the abstracts), and $\gtrsim 1,400,000$
citations since 1990. As an observatory, the Hirsch index of the Hubble Space
Telescope is $\rm h_{HST} \simeq 337$! We argue that it is too soon to
dismantle HST and/or turn off its most productive instruments and operational
modes.

%%% FIGURE 1 %%%
\ve

\vspace*{-0.800cm}
\begin{figure*}[!hptb]
%%% \n\cl{
\begin{minipage}[b]{1.0000\txw}
\hspace*{-0.250cm}
 \begin{minipage}[b]{0.5200\txw}
  \includegraphics[width=0.5150\txw,angle=0]{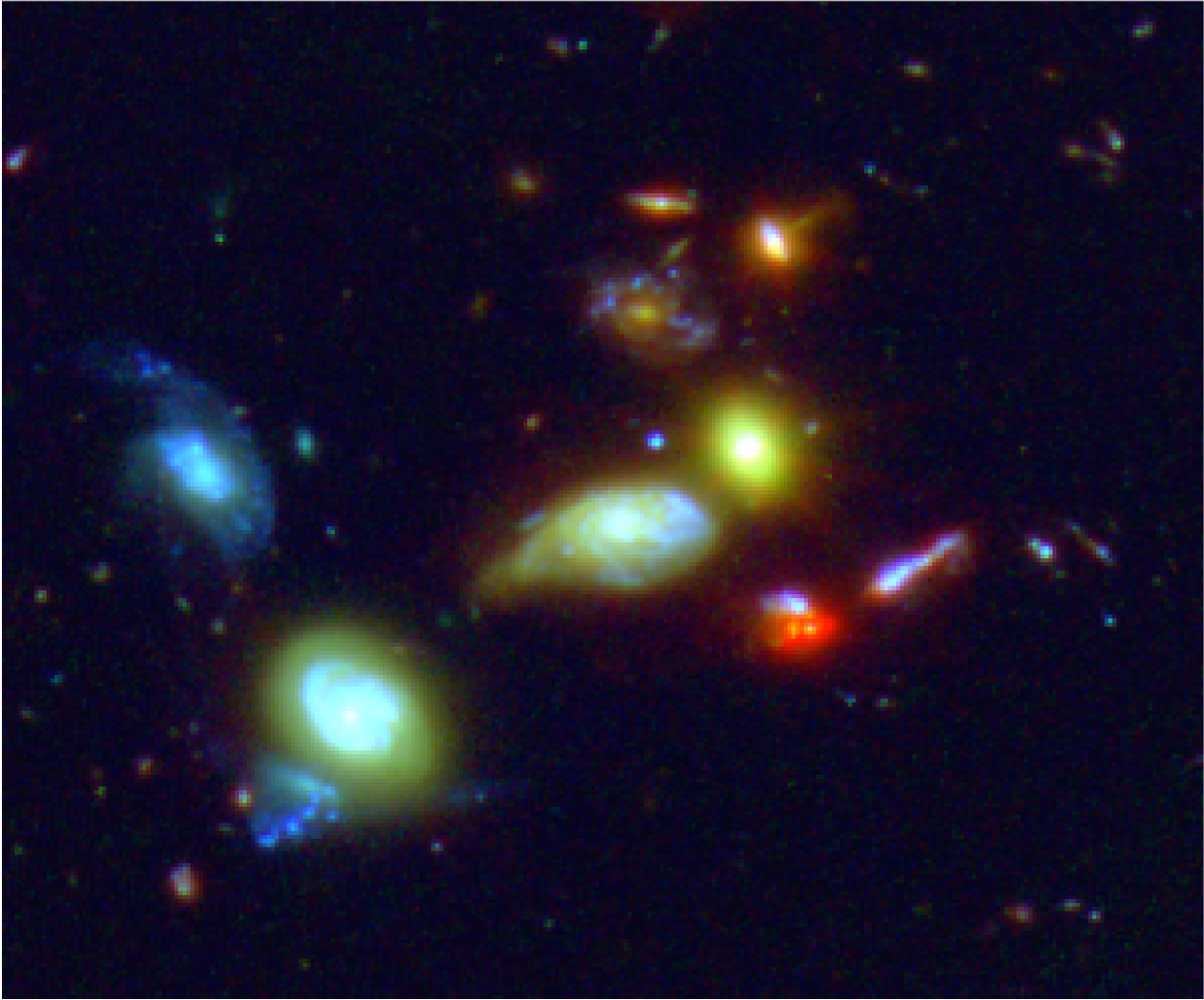}
 \end{minipage}
%%% \hfill
 \hspace*{-0.300cm}
 \begin{minipage}[t]{0.4800\txw}
 \vspace*{-8.400cm}
  \n\includegraphics[width=0.4000\txw,angle=-90]{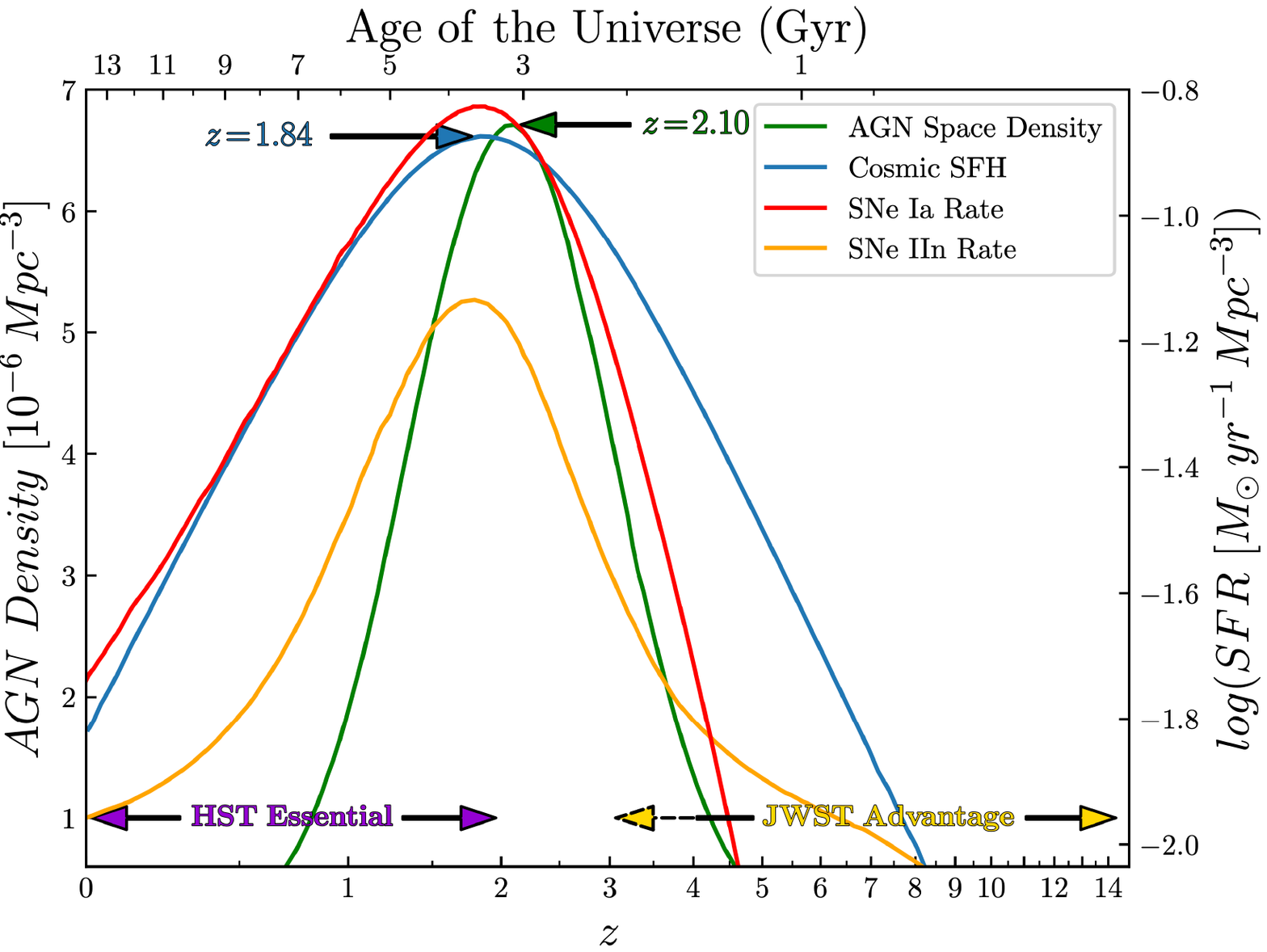}
  \hspace*{+0.500cm}
  \vspace*{-0.400cm}
  \n\includegraphics[width=0.4270\txw,angle=-0]{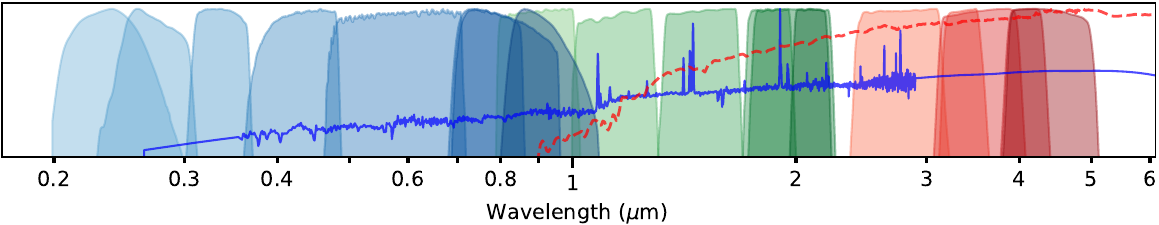}
 \end{minipage}
\end{minipage}
\vspace*{-0.200cm}
\caption{%%% \small %%% \footnotesize %%% \scriptsize 
{\bf Fig 1a. [Left]:}\ A small area form the Hubble UltraDeep Field (HUDF) 
containing galaxies with a wide variety of morphologies and star-formation
histories, illustrating the power of combining 361.3 hours of HST UV--optical
images with 52.7 hours of JWST NIRCam images, rendered in full panchromatic
RGB (full resolution HUDF images are on: 
\url{https://zenodo.org/records/15041742}\ or\ 
\url{https://baas.aas.org/pub/2025i009/release/2}). 
{\bf Fig. 1b. [Top Right]:}\ The cosmic AGN space density, cosmic star-formation
rate, and SNe Type Ia+IIn rates vs. redshift (see \S 3). The peaks of these
curves are indicated and coincide approximately at redshift $z \simeq 2$,
suggesting an underlying relation causing the co-evolution of SMBH masses and
galaxy (spheroid) stellar mass build-up. The galaxy assembly and SMBH epoch of
the last 10 billion years at $z \lesssim 2$ where the HST-unique UV--optical
images are essential are indicated by the purple arrow, while the epoch of the
first 2 billion years at $z \gtrsim 3$ where JWST has its main advantage is
indicated by the orange arrow (the vertical scales have arbitrary
normalization, so that the four curves peak at similar amplitudes). 
{\bf Fig. 1c. [Bottom Right]:}\ Throughput curves of the HST (blue) and JWST
(green and red) filters used to render the full complement of available HUDF
data in Fig. 1a. A star-forming (blue) and an old (red) elliptical galaxy SED
are overlaid at the peak redshift $z \simeq 1.9$ of the cosmic SFH. {\bf
Detecting and monitoring the rest-frame UV--blue radiation from unobscured hot
stellar populations and unobscured accretion disks (weak AGN) is
$\boldsymbol{the}$ unique capability of HST. With its unique resolution and
sensitivity in the UV--optical, long term monitoring by HST is the only way to
capture the faintest SNe and AGN variability signals in significant numbers at
this crucial epoch in cosmic time.}
}
\label{fig:fig1}
\end{figure*}

\vspace*{-0.200cm}
\n\cl{2. WHAT HST HAS DONE SETTING THE STAGE FOR JWST, EUCLID AND ROMAN}

\sn The topic of of galaxy assembly has been one of the main areas where HST
has made significant breakthroughs due to its superb spatial resolution and its
stable dark sky values. The number of HST papers on galaxy assembly has been
too numerous to review here, but we will mention some of the main deep field
work done with HST. Hubble's first Deep Field was the Hubble Deep Field (HDF) for a total of 150
orbits in HST's Continuous Viewing Zone \citep[CVZ;][]{Williams1996}, which was
inspired amongst others by earlier medium-deep field work with the refurbished
telescope since 1994. These HST surveys showed that galaxy assembly started
from smaller sub-galactic clumps that over cosmic time grew into the well known
elliptical and spiral galaxies that we observe today \citep[\eg][and references
therein]{Driver1998}. The Hubble UltraDeep Field (HUDF) and its surrounding
GOODS-South area was carried out in 2003--2004 \citep[][see also Fig. 1a
here]{Beckwith2006, Giavalisco2004}. After the successful installation of the
WFC3 in May 2009, the HUDF, GOODS-South and other fields were followed-up in
the UV and near-IR, resulting in surveys like the WFC3 ERS
\citep[][]{Windhorst2011}, the UV-UDF \citep[][]{Teplitz2013, Rafelski2015},
the HDUV \citep[][]{Oesch2018}, the UDF12 \citep{Koekemoer2013} and the eXtreme Deep Field (XDF)
\citep[][]{Illingworth2013}, and references therein. To help average over the
effects of sample variance these surveys were expanded over more and larger
areas through surveys like CANDELS \citep[][]{Grogin2011, Koekemoer2011} and
COSMOS \citep[][and references therein]{Scoville2007, Kartaltepe2007}. HST's
unique UV--optical performance has thus spectacularly contributed to our
understanding of galaxy assembly and the Cosmic Star Formation History (CSFH).
We now know that cosmic star-formation started at redshifts $z \gtrsim 10$ when
the universe was less than 500 Myr old, but that the CSFH did not peak until $z
\simeq 1.9$ \citep[\ie about 10 Gyr ago;][see Fig. 1b here]{Madau2014} and has
steadily declined since that time \citep[][]{Lilly1996}. Fig. 1b shows that
approximately the same is also true for AGN accretion disks around SMBHs, as
these grew over cosmic time approximately in sync with galaxy bulge growth
\cite[][]{Magorrian1998, Gebhardt2000, Kormendy_2013}, implying that AGN
feedback played a critical role in massive galaxy assembly. Fig. 1c shows the
throughput curves of the HST (blue) and JWST (green+red) filters, to illustrate
the full complement of available HUDF data, with a star-forming (blue) and old
(red) elliptical galaxy SED overlaid at the peak redshift $z \simeq 1.9$ of the
cosmic SFH. {\bf In conclusion, Fig. 1 shows that at least half of all stars in
the universe formed in the era where HST provides its unique rest-frame UV view
of unobscured young, massive stars that trace star-formation, as well as of 
unobscured AGN accretion disks around growing supermassive black holes. Both
HST and JWST sample essential and very complementary phases of galaxy assembly
that thus must continue to be studied with both telescopes!}

\bn\cl {3. WHY HST IS FUNDAMENTALLY COMPLEMENTARY TO JWST AND MUST NOT BE 
DISMANTLED} 

\sn By rendering a subset of the 556.3 hours of available HST images in 12
filters of the HUDF \citep[][]{Beckwith2006, Teplitz2013, Rafelski2015} in an
appropriate mix of colors following the prescriptions of \cite{Windhorst2011}
and \cite{Coe2015}, we illustrate the unique capabilities of HST for galaxy
science emphasizing the rest-frame UV--optical wavelength range. We then
contrast this with the 52.7 hours of publicly available JWST NIRCam images in 8
filters of the same HUDF area from the JADES project \citep[][]{Rieke2023b,
Eisenstein2026}, where we multiplied the actual JWST observing hours by 2 to
represent the total exposure time in each of the simultaneously-exposed NIRCam
SW and LW filter pairs. We render these NIRCam filters in redder colors to
visually illustrate JWST's unique near-IR capabilities to detect older stellar
populations at higher redshifts, as well as very dusty stellar populations and
Active Galactic Nuclei (AGN). Fig. 1a is approximately true color at
$z_{\rm med} \simeq 1.9$, and illustrates that HST uniquely probes (unobscured) young,
hot, massive stars in galaxies, while JWST reveals more advanced stages of
older stellar populations, including relatively short-lived phases where
galaxies produce and shed dust from intense star-formation, and the very rare
high redshift objects ($z \gtrsim 10$) not accessible by HST \citep[see][for
the full high-res PDF files]{Windhorst2025}. Panchromatic studies of the
Integrated Galaxy Light (IGL; see also \S 4b) have shown that {\it
approximately equal parts} of the energy from cosmic star-formation plus cosmic
AGN growth comes from these populations that are: a) largely unobscured by dust
in the rest-frame UV--optical; and b) significantly obscured by dust in the
rest-frame UV--optical, and therefore best detected from near-IR to millimeter
wavelengths \citep[][and references therein]{Driver2016b, Andrews2018,
Koushan2021}. Hence, as we will detail below, {\bf HST alone is as important to
image these populations as facilities like, $\boldsymbol{e.g.}$, JWST, Euclid, Roman, Spitzer,
WISE, Herschel and ALMA are combined.} HST unique UV--optical imaging
capabilities adds the unobscured restframe UV--blue light that samples the more
recent half of the CSFH in the last 10 billion years since $z \lesssim 2$, when
compared to the JWST restframe optical--near-IR light that samples the older
and/or more dusty stellar populations, and the very few rare objects at $z
\gtrsim 10$ seen in the first 500 Myr.

\bn\cl {4. RECENT RESULTS THAT USE ARCHIVAL OR NEW HST IMAGES, ESPECIALLY IN
THE UV--OPTICAL} 

\si {\it 4a. Lyman Continuum Escape studies at $z\simeq~$2--3.5 to find the
main population(s) maintaining Cosmic Reionization:}\ The experiment of where,
when and how LyC radiation escapes from reionizing sources --- (dwarf) galaxies 
and/or AGN --- can be done in several ways. One is to measure LyC flux at less
than 912 \AA\ by spectroscopy with HST STIS or COS, or by imaging with STIS/ACS
MAMA detectors. This the focus of other white papers. With UV-optimized CCDs in
WFC3/UVIS, we can measure LyC escape fraction for $z>2$ using F225W, F275W and
F336W filters, and for $z\lesssim3.5$ where the Lyman forest allows fewer
clear lines-of-sights (LOS) at the higher redshifts to permit LyC radiation to
travel through. This HST LyC work has been pioneered with WFC3/UVIS in the WFC3
ERS field \citep[][]{Smith2018, Smith2020}, in the HDUV field
\citep[][]{Naidu2017, Oesch2018}, and in the UVCANDELS survey
\citep[][]{Smith2024, Wang2025}. The evidence available thus far suggests that
weak AGN are more successful in producing escaping LyC radiation than (dwarf)
galaxies on average \citep[][]{Smith2020, Smith2024}, which is also supported
by recent theoretical work \citep[][]{Madau2015, Madau2024}. However, the currently
available statistics on directly detected escaping LyC radiation remain small,
and the measurements are difficult to accomplish, and so {\bf this work needs
to continue for as long as possible with HST WFC3/UVIS} in the rest-frame UV to
complete the details of where, when and how LyC escapes and completed the
process of cosmic reionization.

\si {\it 4b. Multi-Decade Monitoring of the Panchromatic HST Background to
constrain Zodiacal and Diffuse Light:}\ What is often overlooked is that
precisely the combination of HST's very dark sky surface brightness (sky-SB)
environment {\it and} its ability to perform precision photometry on timescales
of minutes to decades offers significantly new parameter space where unexpected
discoveries can be made. At 0.8--1.6 \mum, the HST sky-SB in occultation is
between $\sim~$10--1000$\times$ darker than what can be obtained from the
ground with broad-band observations. Using WFPC2 (since December 1993), ACS
(since March 2002), and WFC3 (since May 2009), well over 1.7 million images have
been taken with HST, many of them in random areas of sky. Details of this work
--- amongst others done through multi-year HST Archival Legacy project
``SKYSURF'' --- can be found in \cite{Windhorst2022, Carter2026,
Tompkins2026}, who compiled approximately 249,000 independent HST exposures
plus 891,000 WFC3/IR on-the-ramp integrations to monitor the sky-SB values from
Low Earth Orbit (LEO) in between the detected discrete objects, most of which,
of course, are faint galaxies that make up the IGL, and Galactic stars. These
HST sky-SB measurements are on an {\it absolute flux scale} with a long term
precision of $\lesssim~$2.7--4\%, including all known sources of systematics
for WFC3/IR, ACS/WFC, WFC3/UVIS and WFPC2, respectively. When compared to the
best available Zodiacal model of \cite{Kelsall1998}, \cite{Carleton2022} finds 
a residual amount of Diffuse Light (DL) that is present in the HST images of the
order of $\lesssim~$29--40 \nWsqmsr\ at 1.25--1.6 \mum. \cite{OBrien_2023,
OBrien_2026} and \cite{McIntyre2025} extended this work down to the WFC3/UVIS
wavelengths of 0.2 \mum, and found residual DL levels to be less than
$\lesssim~$21--32 \nWsqmsr, with lower bounds of $\gtrsim~$2--20 \nWsqmsr\ at
1.25--1.6 \mum, respectively. These are the best estimates available for the
DL, and these results provide an interesting perspective compared to recent
work with the New Horizons (NH) spacecraft from beyond Pluto's orbit, which
finds DL levels $\lesssim~$3.0 \nWsqmsr\ \citep{Postman2024}. The best Zodiacal
Light models, which have been around for over 28 years \citep{Kelsall1998}, may
thus be missing a very dim (spherical or spheroidal) level of Diffuse Light in
the inner solar system that HST clearly detects. \cite{OBrien_2026} suggests
that this extra diffuse light might come from icy dust particles left in the
inner solar system by comets on their way around the sun, and updated the
\citet{Kelsall1998} Zodiacal Light model accordingly using 20+ years of HST
images at 0.2--1.6 \mum. It now seems unlikely that this Diffuse Light comes
predominantly from cosmological distances. This work with HST can now be done
thanks to the stable zeropoints to (well) within \cle~1--2\% over decades and
the resulting well calibrated HST images. Archival HST work is most difficult
at its UV--blue wavelengths, hence {\bf continued HST monitoring of the sky-SB
at 0.2--0.9~$\boldsymbol{\mu}$m will add valuable and tighter constraints to any DL that may
come from outside the solar system.}

\bn\cl {5. ONGOING AND FUTURE UV--OPTICAL HST WORK COMPLEMENTING JWST, EUCLID,
AND ROMAN}

\si {\it 5.a UV--Optical Variability of Weak AGN in the context of Galaxy
Assembly:}\ HST remains essential to observe the extremely faint rest-frame
UV--optical signals that can be associated with unobscured time-variable AGN.
Previous studies \citep{Cohen_2006, Sarajedini_2003, Villforth_2010,
Pouliasis_2019, Zhong_2022, OBrien_2024, Hayes2024} were able to identify
variability as faint as $\sim0.2$ mag. For AGN largely unobscured by dust from
the surrounding torus or galaxy --- \ie\ those AGN whose accretion disk and its
outflow cone shine mostly in our direction --- detection of AGN variability in
the rest-frame UV--optical is optimal, as variability has been shown to
increase at the shorter wavelengths \citep[][]{Paltani_1994, diClemente_1996,
Helfand_2001}. Hence, HST is {\it the} platform of choice to monitor weak AGN
variability at $z \lesssim 2$ --- where half of the cosmic AGN growth takes
place (Fig. 1) --- and constrain their accretion rates (for SMBH
accretion disks more significantly obscured by dust, JWST is needed of course
to identify the central AGN component \citep[\eg][]{Ortiz2024, Bowling2026}).
HST's resolution is essential to separate the rest-frame UV accretion disks of
AGN from the host galaxy for resolved objects. HST's F435W and F606W filters 
provide the optimal combination of depth (as faint as $\rm AB \sim 28.0$ mag)
and increased variability amplitude. Longer wavelengths would cause the
variability to appear over longer time-scales, while shorter wavelength filters
have relatively lower sensitivity.

\si {\it 5.b. Long-Term HST UV--Optical Monitoring of SNe Ia as Standard 
Candles:}\ A major issue in using luminosity distances from SNe Ia is a possible
intrinsic evolution of SN properties with the age and metallicity of the
progenitor stars. The existing samples SNe Ia have shown residual correlations
with host-galaxy properties, perhaps reflecting uncorrected age, metallicity,
and dust differences among progenitors \citep[][]{MSmith2020, Nicolas2021,
Brout2021, Kelsey2021, Briday2022}. Lower metallicities at higher redshift
implies that the progenitor stars are less likely to lose mass through stellar
winds. As a result, the average mass of the white dwarfs at high-$z$ is
expected to be larger than their $z \simeq 0$ counterparts. These systematic
differences are expected to be the most significant at redshifts
$z \simeq~$1.5--3.0 \citep[][see also Fig. 1 here]{Riess2006}. Optical HST
imaging in the blue is perfectly suited to identify these explosions and the
progenitor host galaxy in great detail, enabling the study of environmental
effects on SNe Ia's and their progenitor white dwarfs. Evidence for a
dependence of SNe Ia luminosities on progenitor metallicities, age, or other
environmental effects would have profound implications on the current
cosmological paradigm. 

\si {\it 5.c. HST Studies of Individual Stars at Cosmological Distances via
Cluster Caustic Transits:} One of the most stunning --- and rather unforeseen
--- discoveries of HST was the detection of individual stars at cosmological
distances through cluster caustic transits \citep{Welch2022a}, as predicted by 
\citep{MiraldaEscude1991}. Magnifications of many factors of 1,000--10,000 are
in principle possible, as the magnification factor scales as $\mu \propto
(10-20)/\sqrt{d}$, where $d$ is the distance of the star (in arcsec) to the
caustic. This could temporarily boost the brightness of a very compact object
by $\mu \gtrsim~$7.5--10 mag on timescales of months, since stars at
cosmological distances will have angular diameters of order $\sim
10^{-11}$--10$^{-10}$ arcsec, and the clusters can have lateral velocities in
the cosmic flow of $v \simeq 1,000$ \kms\ \citep[][]{Kelly2018, Windhorst2018}.
Cluster microlensing also yields interesting constraints to the Dark Matter
(DM) nature and content of the lensing cluster's ICM \citep[][]{Diego2024a,
Diego2024b, Broadhurst2025, Pozo2026}. While originally thought to be the
exclusive domain of detections by JWST at $z \gtrsim~$6--7, a number of caustic
transits have been found with HST, where its blue performance is essential for
their discovery and proper interpretation. Most stellar caustic transits
discovered so far have the SEDs or spectra of double stars/compact clusters.
The sample thus far consists of luminous Blue Super Giants (BSGs) best detected
by HST at intermediate redshifts, and of Red Super Giants (RSGs) best detected
with JWST. Finally, stunning numbers of over 40--100 caustic transits (!) were
discovered including essential HST data in the Dragon's arc at $z=0.725$ behind
the cluster Abell 370 at $z=0.37$ \citep[][]{Fudamoto2025, Broadhurst2025,
Palencia2026}. While JWST provided the second epoch for this unbelievable
result, HST provided the essential first-epoch deep imaging, as well as the
leverage to separate BSGs from RSGs. {\bf These results are fundamental in that
they are starting to allow us to directly constrain the (top end of the)
Initial Mass Function (IMF) at cosmological distances by counting individual
magnified stars. HST must continue the optical blue part of this work.}

\bn\cl {6. KEEPING HST's UNIQUE CAPABILITIES ALIVE IN THE CONTEXT OF FUTURE
SPACE MISSIONS} 

\sn It is also critical to keep HST's unique capabilities alive for as long as
possible in the context of future space missions: 

\si \bul (1) The ESA Euclid mission \citep[][``Euclid'']{Mellier2025}, launched in July
2023, is a 1.2 m telescope with a Wide Field Imager (WFI) that is covering
$\sim 14,000$ \degsq\ in 4 broad-band filters spanning 0.6--2~\mum\ to $\rm AB
\lesssim 26.7~$mag. WFI has one visual filter that spans a very wide range of
5500--9000 \AA\ designed for weak lensing. Euclid also provides spectroscopic
redshifts for brighter objects using two near-IR grisms.

\si \bul (2) The Nancy Grace Roman Space Telescope
\citep[][``Roman'']{Akeson2019}, scheduled for launch later in 2026, is a 2.4 m
telescope with a Wide Field Imager. Its community surveys will map many 100's
of \degsq\ to $\rm AB \lesssim~$28--29 mag in 6 filters between 0.6--2~\mum,
although only its bluest filter (R062) overlaps ``HST-unique'' parameter space,
in that JWST does not have an equivalent filter. Both Euclid and Roman aim to
constrain the cosmological parameters through a combination of Baryon Acoustic
Oscillations (BAO), weak lensing, and/or high redshift Supernovae, but lack the
very large suite of HST-unique filters between 0.2--0.6 \mum, so they cannot
replace HST's role in Fig. 1c. Moreover, in the narrow wavelength range where
they do overlap with HST, the detector sampling (0\arcspt 11/pixel for Roman; 
\cite{Akeson2019}, and 0\arcspt 298/pixel for Euclid; \cite{Mellier2024}).
Hence, the effective angular resolution attained by these missions is $\sim$
2.8--7.5$\times$ worse than HST's WFC3/UVIS \citep[which has 0\arcspt
039/pixel; \eg][]{Windhorst2011}, so that their sensitivity per unit time to
the faintest point sources $\sim7.8$--$57\times$ worse than HST's. {\bf Neither
Roman nor Euclid can replace HST's unique UV--blue capabilities.}

\si \bul (3) Lazuli is a 3 m optical space telescope designed for monitoring
of fast evolving transients, SNe and exoplanets with rapid IFU follow-up of
single targets, to be launched this decade \citep{Roy_2026}. While very
powerful for these purposes, its filter set is broad-band SDSS $ugriz$, and
lacks the vacuum UV and extensive medium+narrow-band filter sets of HST WFC3
that are essential for detailed physical studies of sizable astrophysical
samples. 

\si \bul (4) The Chinese Space Station Telescope (CSST or ``Xuntian'') is a 2 m
telescope also to be launched in late 2027 \citep[][]{Cao2022}. The aim for
CSST is to cover $\sim 17,500$ \degsq\ in 7 broad- and 9 medium-band filters
between 0.25--1.0 \mum\ to $\rm AB \lesssim~$25--26 mag \citep[see Table 1
of][]{Gong2019} with science goals like weak lensing and galaxy clustering.
Xuntian will have PSF-widths of $\sim~$0\arcspt 20--0\arcspt 24 FWHM, so that
the CSST PSF will cover a \cge~25$\times$ larger area than the HST UV--optical
PSF area, resulting in a similar loss in depth. Hence, the CSST will {\it not}
enable the high resolution studies like those illustrated in Fig. 1a for the
fainter and smallest galaxies. A significant fraction of faint galaxies would
simply remain unresolved at 0\arcspt 24 resolution to $\rm AB \lesssim 27~$mag
in the UV--optical (see, \eg\ \cite{Kramer2022}, Fig. 10a of 
\cite{Windhorst2011} and Fig. 6a of \cite{Windhorst2023}). {\bf NASA cannot
afford to dismantle the unique HST because we may some day have CSST or Lazuli.}

\bn\cl{7. CONCLUSIONS} 

\sn In conclusion, HST is essential to image the unobscured cosmic
star-formation and the growth of AGN accretion disks in the last 10 billion
years since $z \lesssim 2$, and JWST is essential to image the start and
ramp-up of cosmic star-formation and AGN growth in the first few billion years.
Roman will find rare red--near-IR targets where only HST can provide the
UV--optical for full characterization. {\bf In summary, using full cost
accounting, the HST flagship mission will have costed over 20 B\$ since the
early 1970's (including all 6 Shuttle missions), while the JWST flagship
mission will have costed over 10 B\$ since the mid 1990's, and Roman over 3 B\$
since 2010. Therefore, to get full return on our investment in JWST and Roman,
HST needs to continue its unique and strong complementary UV--optical role for 
as long as possible.} It is also essential to keep Hubble's mission funding
strong, not only to keep the HST-unique instruments running for as long as
possible, but also to maintain viable grant funding for HST observers, which
will train the next generation in HST work that prepares them well to work on
future space missions like the 6.5 m Habitable Worlds Observatory \citep[HWO,
\eg][]{Feinberg2026}. Indeed, HST's truly unique high-resolution vacuum UV--blue
performance will be {\it the} perfect training ground for HWO.

\ve %%%%

%%%%%%%%%%%%%%%%%%%%%%%%%%%%%%%%% ACKS and REFS %%%%%%%%%%%%%%%%%%%%%%%%%%%%%%%%%%%

\sn {\bf Acknowledgments:} We thank Dr. Christopher Willmer for his help
clarifying the effective HUDF exposure times in the public JADES data. We
acknowledge support from HST grants that enabled us to pursue our more recent
Hubble work, such as HST-AR-13877, HST-AR-14591, HST-GO-15278, HST-GO-15647,
HST-AR-15810, HST-GO-16252, HST-GO-16605, HST-GO-16621, and HST-GO-16793,
provided by NASA through the Space Telescope Science Institute, which is
operated by the Association of Universities for Research in Astronomy, Inc.,
under NASA contract NAS 5-26555. RAW, SHC, and RAJ acknowledge support from
NASA JWST Interdisciplinary Scientist grants NAG5-12460, NNX14AN10G and
80NSSC18K0200 from GSFC. We also acknowledge the indigenous peoples of Arizona,
including the Akimel O'odham (Pima) and Pee Posh (Maricopa) Indian Communities,
whose care and keeping of the land has enabled us to be at ASU's Tempe campus
in the Salt River Valley, where much of our work was conducted.

\bibliographystyle{aasjournal}
\bibliography{references_pearls_overview_v2}{}

\end{document}